\documentclass[a4paper,11pt]{article}
\pdfoutput=1 

\usepackage{jinstpub} 

\usepackage{braket}
\usepackage{subcaption}
\usepackage{textcomp}
\usepackage{amssymb}

\usepackage{lineno}

\title{Novel imaging technique for $\upalpha$-particles using a fast optical camera}

\author[a]{Gabriele D'Amen\footnote{Corresponding author.},}
\author[a]{Michael Keach,}
\author[a]{Andrei Nomerotski,}
\author[b,c]{Peter Svihra,}
\author[a]{Alessandro Tricoli}

\affiliation[a]{Brookhaven National Laboratory, Upton NY 11973, USA}
\affiliation[b]{Department of Physics, Faculty of Nuclear Sciences and Physical Engineering, Czech Technical University, Prague 115 19, Czech Republic}
\affiliation[c]{Department of Physics and Astronomy, School of Natural Sciences, 
University of Manchester, Manchester M13 9PL, United Kingdom}

\emailAdd{gdamen@bnl.gov}

\abstract{A new imaging technique for $\upalpha$-particles using a fast optical camera focused on a thin scintillator is presented. 
As $\upalpha$-particles interact in a thin layer of LYSO fast scintillator, they produce a localized flash of light. The light is collected with a lens to an intensified optical camera, Tpx3Cam, with single photon sensitivity and excellent spatial \& temporal resolutions. The interactions of photons with the camera is reconstructed by means of a custom algorithm, capable of discriminating single photons using time and spatial information.}


\begin{document}
\maketitle
\flushbottom

\section{Optical detection of $\upalpha$-particles}
\label{sec:intro}

Though multiple ways to detect $\upalpha$-particles have been developed over the years, interest in detecting radiation in real-time with improved sensitivity and spatial \& temporal resolutions recently arose. This need for fast $\upalpha$-imaging cameras is mostly driven by nuclear medicine \cite{Back2010, Miller2018}, nonproliferation, and other security applications \cite{Morishita2014}.

Due to the large energy deposition of $\upalpha$-particles emitted by typical radioactive sources, many approaches can be used. Alphas can be detected in solid state pixelated sensors using a variety of possible readouts, such as the Timepix chip \cite{Jakbek2009}. Optical schemes can also be applied to the $\upalpha$ detection like in the iQID camera, where a scintillator is coupled to an image intensifier \cite{Miller2014, Miller2015}.
In the optical regime, good coordinate resolution was obtained by employing a position sensitive scintillation detector based on bundled LYSO crystal fibers, coupled to micro-pixel avalanche photodiodes and read out by a CCD \cite{Yamamoto1997} or, later, with an EMCCD \cite{Yamamoto2018}. 
Recently, a fast single-photon sensitive optical camera, Tpx3Cam, was employed for the registration of $\upalpha$-particles in a gaseous TPC \cite{Roberts2019} for studies related to three-dimensional optical readout of a dual-phase liquid Argon-TPC (time projection chamber) \cite{Mavrokoridis2015, Hollywood2020}. 

We describe below another experimental implementation of the optical registration by employing a thin LYSO scintillator where light flashes are produced by $\upalpha$-particles from an $^{241}$Am radioactive source. Photons from these flashes are collected with a lens onto an intensifier coupled to the Tpx3Cam optical camera, with excellent temporal and spatial resolution. This setup is illustrated in Figure \ref{fig:setup}.
The proposed optical approach provides a device with nanosecond scale time resolution and good position resolution.
Though the light collection at long distances could pose a problem, clear advantages of the technique are the remote location of the only active detection element, i.e. the camera, and the flexibility of the light collection schemes, which can include mirrors to place the camera outside of the beam.

The material is organized as follows. Section \ref{sec:setup} describes the experimental setup, including the fast camera used for the measurements. Section \ref{sec:analysis} provides a detailed account of the analysis algorithms and describes main results of the study. Section \ref{sec:discussion} discusses advantages of the optical technique and Section \ref{sec:conclusions} draws conclusions.

\section{Experimental Setup}
\label{sec:setup}

\begin{figure}[htbp]
    \centering
    \includegraphics[width=0.8\linewidth]{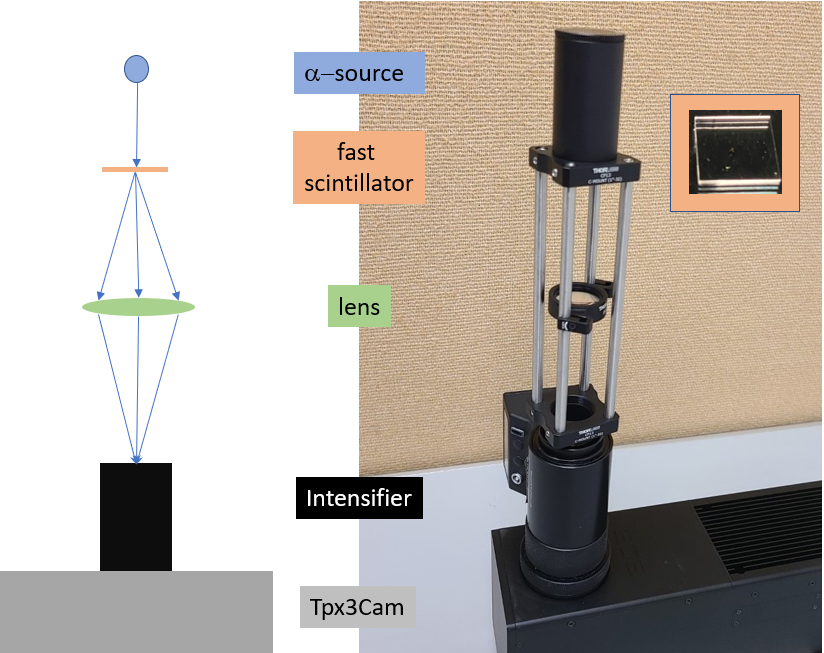}
    \caption{Left: Optical detection scheme of $\upalpha$-particles. An $\upalpha$ interacts in a thin layer of fast scintillator producing a flash of light. The light is collected with a lens on to an intensified optical camera with single photon sensitivity, and good spatial and temporal resolution. Right: Photographs of the setup, and LYSO scintillator in the top-right insert. }
    \label{fig:setup}
\end{figure}

The experimental setup is presented schematically in Figure \ref{fig:setup}: on top, a 6 kBq americium-241 $\upalpha$-source was used to illuminate a thin LYSO scintillator with dimensions $10 \times 10 \times 0.5$~mm$^3$, placed at a close distance. The scintillator was positioned at a distance of 10~cm from the input window of the intensifier in the Tpx3Cam and oriented orthogonally to the beam axis, as shown in Figure \ref{fig:setup}. The lens used to focus the scintillator light flashes was positioned on the sliding rods at approximately the midpoint between the scintillator and the intensifier, with the exact location chosen by a focusing procedure. A custom 3D-printed box (not shown in Figure \ref{fig:setup}) provided the light-tight enclosure for the source, scintillator, lens and intensifier.

Imaging of single photons in this experiment was performed using the Tpx3Cam. The Tpx3Cam is a single-photon sensitive camera with time stamping capabilities \cite{timepixcam, tpx3cam, ASI}, which has a data-driven readout with pixel dead time of only about a microsecond
allowing for multi-hit functionality for each pixel independently and bandwidth of up to 80 Mpix/sec.  Additionally, the Tpx3Cam optical sensor has a high quantum efficiency (QE) \cite{Nomerotski2017}. This optical sensor has 256x256 pixels, each with dimensions of 55x55 $\mu$m$^2$, and is bump-bonded to the Timepix3 readout chip, which is an application specific integrated circuit (ASIC) \cite{timepix3}. Incoming signals are processed pixel-by-pixel by the electronics  with a precision on time of arrival (ToA) of 1.56~ns for hits crossing a predefined threshold. Each pixel also contains a memory register, which stores information about the ToA, as well as the time-over-threshold (ToT) of each hit, which is related to the energy deposited in the pixel.

To detect single photon emissions from the scintillator, a cricket with integrated image intensifier is used to convert these single photons to light flashes and project them onto the optical sensor on the camera. The image intensifier is a vacuum device comprised of a photocathode followed with a micro-channel plate (MCP) and fast P47 scintillator. The hi-QE-green photocathode in the intensifier (Photonis \cite{Photonis}) has a QE of about 30\% at 430~nm, which is the maximum of emission spectrum of the LYSO scintillator. The MCP in the intensifier has an improved detection efficiency close to 100\%.
Similar configurations of the intensified Tpx3Cam were used before for characterization of quantum networks \cite{Ianzano2020, Nomerotski2020}, quantum target detection \cite{Yingwen2020, Svihra2020}, single photon counting \cite{sensors2020} and lifetime imaging \cite{Sen2020} studies.

Prior to measurements, the camera underwent a calibration process. During this process, the threshold levels of individual pixels are adjusted, equalizing their effective thresholds to fast light flashes from the intensifier to $600-800$ photons per pixel, depending on the wavelength. Additionally, a small ($\approx0.1\%$) number of hot pixels were masked off to prevent recording of a large number of noise hits.

\section{Data Analysis}
\label{sec:analysis}


At first, a standard post-processing procedure was applied, as developed and optimised for single photon detection, and documented in Refs. \cite{tpx3cam, Ianzano2020}. This method does not take into account the discrimination of multiple photons from a single $\upalpha$-particle signal or the substantial superimposition and blending of hits on the sensor. After a focusing procedure, aimed to minimize the spatial spread of single photons at the intensifier input window, several datasets lasting 100 seconds were recorded.

After being sorted by ToA, hits on the pixels are grouped into "clusters" using a recursive algorithm. Clusters are small collections of hits on pixels adjacent to each other and registered within a time window of a predefined length of 300~ns. 
Since all pixels measure ToA, ToT and position information of the hits independently, a centroiding algorithm can be used to determine the coordinates of single photons by combining these measurements. The $x$ and $y$ coordinates of the impinging single photon is estimated as a weighted average of the sensor hit positions, using the ToT information as a weight. The timing of the photon is assumed to be the ToA of the pixel with the largest ToT in the cluster. The ToA is corrected by accounting for the time-walk, i.e. an effect caused by the variable pixel electronics time response to input signals of different amplitudes \cite{Turecek_2016, tpx3cam}. The magnitude of the correction to ToA is computed by fitting a function of the ToT of the corresponding pixel hit and of the threshold of the detector discriminator. By applying this correction, a 2~ns time resolution (rms) can be achieved for single photons \cite{Ianzano2020}.

\begin{figure}[htbp]
  \centering
  
  \begin{subfigure}{\linewidth}
        \fbox{\includegraphics[width=0.49\linewidth]{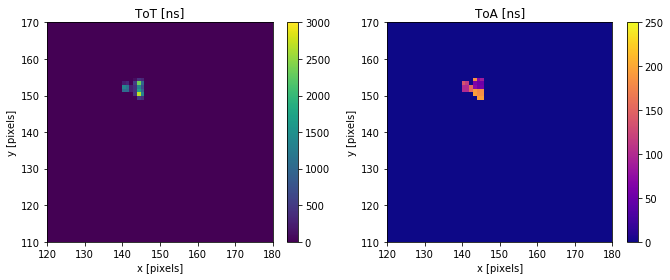}}
        \hspace{0.01\textwidth}
        \fbox{\includegraphics[width=0.49\linewidth]{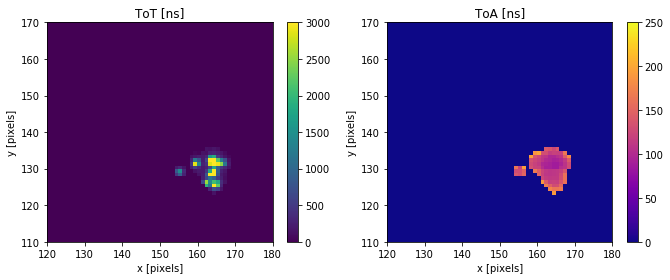}}
    \end{subfigure}
    \newline
    \par
    
    \begin{subfigure}{\linewidth}
        \fbox{\includegraphics[width=0.49\linewidth]{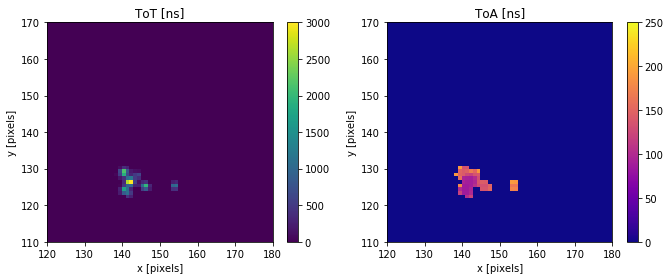}}
        \hspace{0.01\textwidth}
        \fbox{\includegraphics[width=0.49\linewidth]{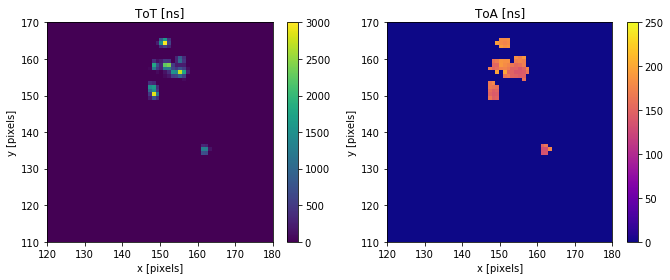}}
    \end{subfigure}
    \newline
    \par
    
    \begin{subfigure}{\linewidth}
        \fbox{\includegraphics[width=0.49\linewidth]{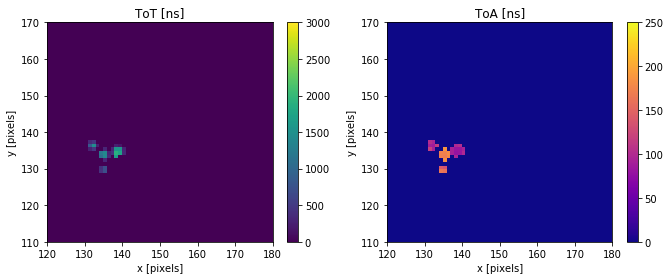}}
        \hspace{0.01\textwidth}
        \fbox{\includegraphics[width=0.49\linewidth]{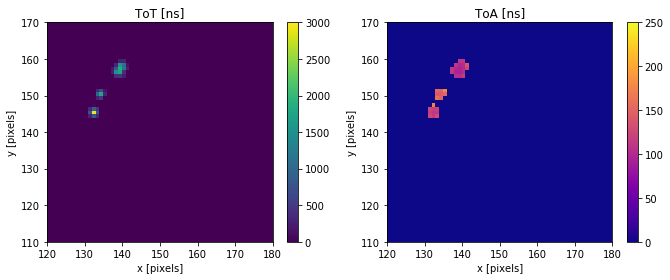}}
    \end{subfigure}
    \newline
    \caption{Six examples of hits induced by $\upalpha$-particles in the selected ROI of $60 \times 60$ pixels. The hits are shown as boxed pairs of heatmaps in ToT representation (left graph in the boxed pair of graphs, ToT in ns) and ToA representation (right graph, ToA in ns from the earliest hit pixel in the shown data). The ToA values are shown before the time-walk correction. }
    \label{fig:example}
\end{figure}

The hits induced by $\upalpha$-particles appear as collections of multiple single photons detected by the camera. Figure \ref{fig:example} shows six examples of hits induced by $\upalpha$-particles, selected in a region of interest (ROI) of $60 \times 60$ pixels corresponding to the $^{241}$Am source spot. 
Here events were purposely selected with a large number of pixels featuring similar values of ToA. Note that good focusing would result in multiple photon hits blending, i.e. the spatial superimposition of multiple photons created in the LYSO by a single alpha, reaching the sensor at the same time. These examples illustrate the richness of the recorded information, which leaves a considerable scope for possible reconstruction algorithms. The provided supplemental materials includes more examples of $\upalpha$-particle hits.

\begin{figure}[htbp]
    \centering
    \includegraphics[width=0.45\linewidth]{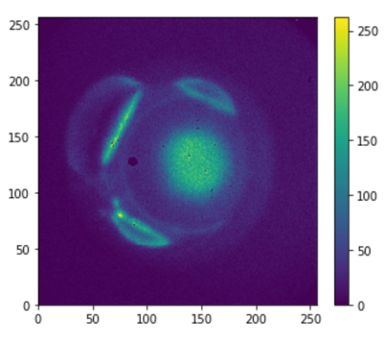}
    \includegraphics[width=0.45\linewidth]{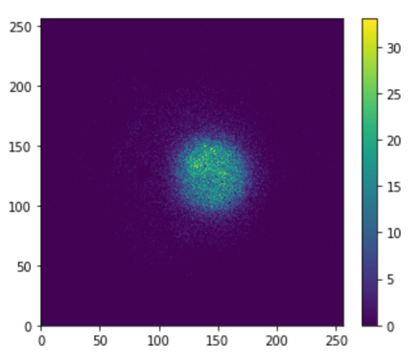}
    \caption{Occupancy of all reconstructed clusters (left) and of the clusters with more than 20 pixels (right) as function of the x-y position in the sensor. The position of radioactive source is centered at (150, 120) where the coordinates are in pixels.}
    \label{fig:occupancy}
\end{figure}

Figure \ref{fig:occupancy} shows the pixel occupancy of all reconstructed clusters and of the clusters with more than 20 pixels. The data is shown for the full $256 \times 256$ pixel area of the sensor. One can clearly see the central spot at (150,120) corresponding to the activated area of the $\upalpha$-source. In the left graph the edges of the $10 \times 10$~mm scintillator plate are visible due to photons reflected inside the scintillator, which exit through the edges and are then reflected towards the camera. The amount of such photons is small and they can be removed by requiring a large number of hit pixels in a cluster, as shown in the right graph where a requirement of more than 20 pixels per cluster was applied.

\begin{figure}[htbp]
    \centering
    \includegraphics[width=0.8\linewidth]{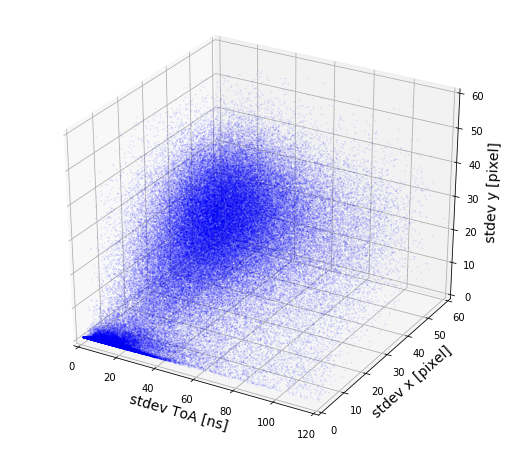}
    \caption{Three-dimensional distribution of standard deviations of $x$ and $y$ coordinates of all pixels in clusters with the time requirement applied as detailed in the text, and the corresponding standard deviation of the time-walk-corrected ToA. 
    }
    \label{fig:timecluster}
\end{figure}

In order to reconstruct clusters originating from multiple photon hits, the requirement of pixels to be adjacent is removed from the clustering algorithm, while it is required that all the pixels in a cluster are within a predefined time window of 300~ns. With this condition, all hit pixels in $\upalpha$-induced events are combined into temporal clusters. 
The cluster size can be characterized by spatial and temporal variances of the pixels in a cluster. 
Figure \ref{fig:timecluster} shows a three-dimensional distribution of standard deviations for $x$ and $y$ coordinates of all pixels in such clusters along with the corresponding standard deviation of ToA. Large spreads in hits distribution indicate the presence of cases when multiple photons were detected. These photons can be separated spatially by the geometry as well as reflections and focusing effects, and temporally by accounting for the LYSO emission decay time. 
One can identify two classes of events in Figure \ref{fig:timecluster}, one with small spatial and temporal variance near zero and one whose spread in all dimensions is considerably larger, corresponding to events with large numbers of photons, as shown in the event selection of Figure \ref{fig:example}. It is hypothesized that the former are events from the 59~keV x-ray emission that occurs in the $^{241}$Am isotope at the same rate as the $\upalpha$-particle emission. Since there is no requirement for the hits to originate in the source spot, 
the additional scatter of the photon clusters is due to the photons exiting on the sides of the scintillator after reflections. We note that while increasing the spatial variation those photons do not improve the spatial resolution if used in the centroiding algorithm. Due to the much smaller energy deposition, the x-rays produce primarily single photon clusters which, being compact, are grouped near zero in the three-dimensional distribution.




\begin{figure}[htb]
    \centering
    \includegraphics[width=.49\linewidth]{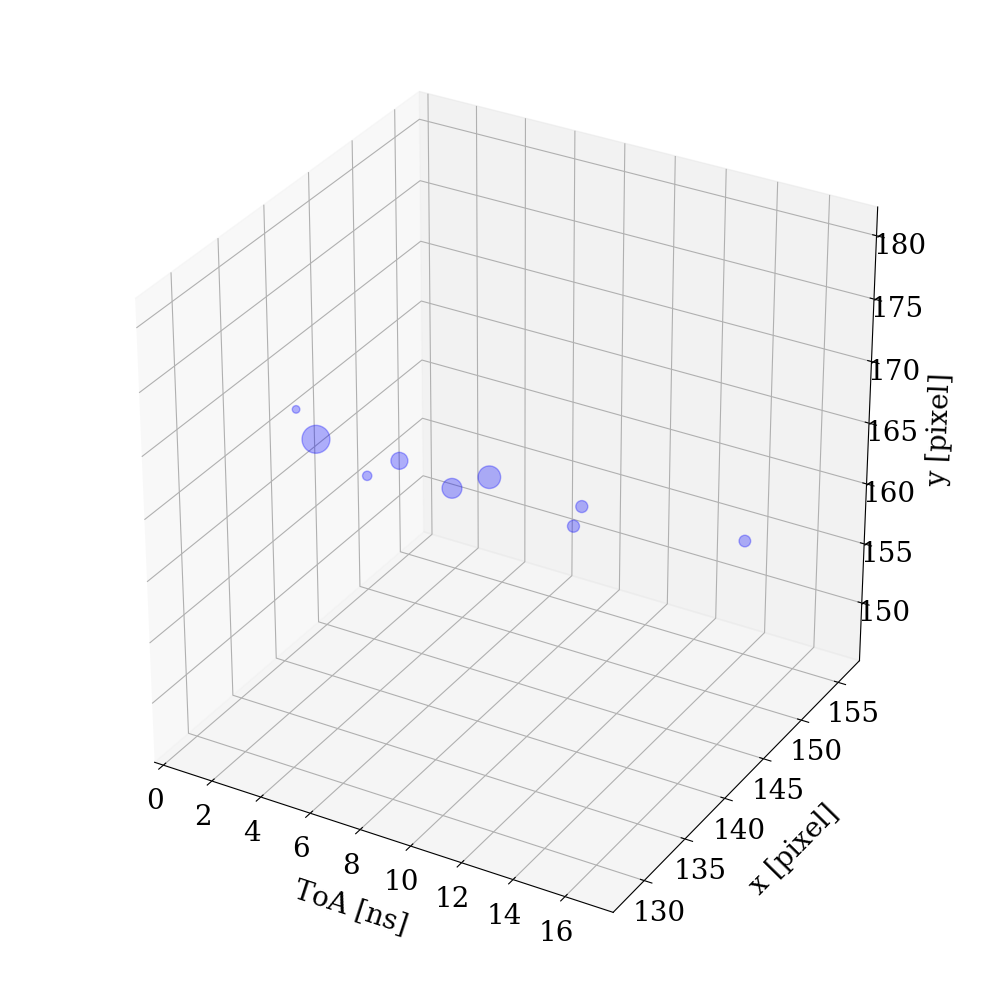}
    \includegraphics[width=.49\linewidth]{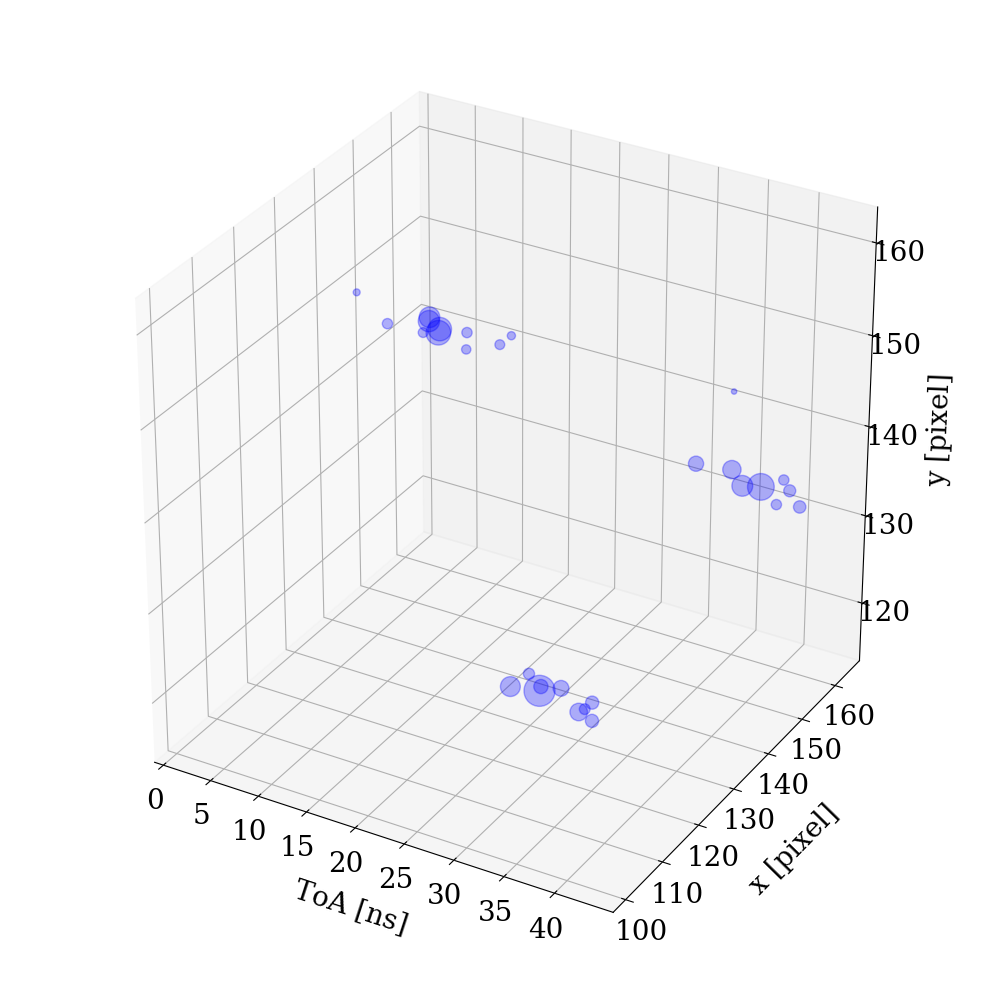}
    \includegraphics[width=.49\linewidth]{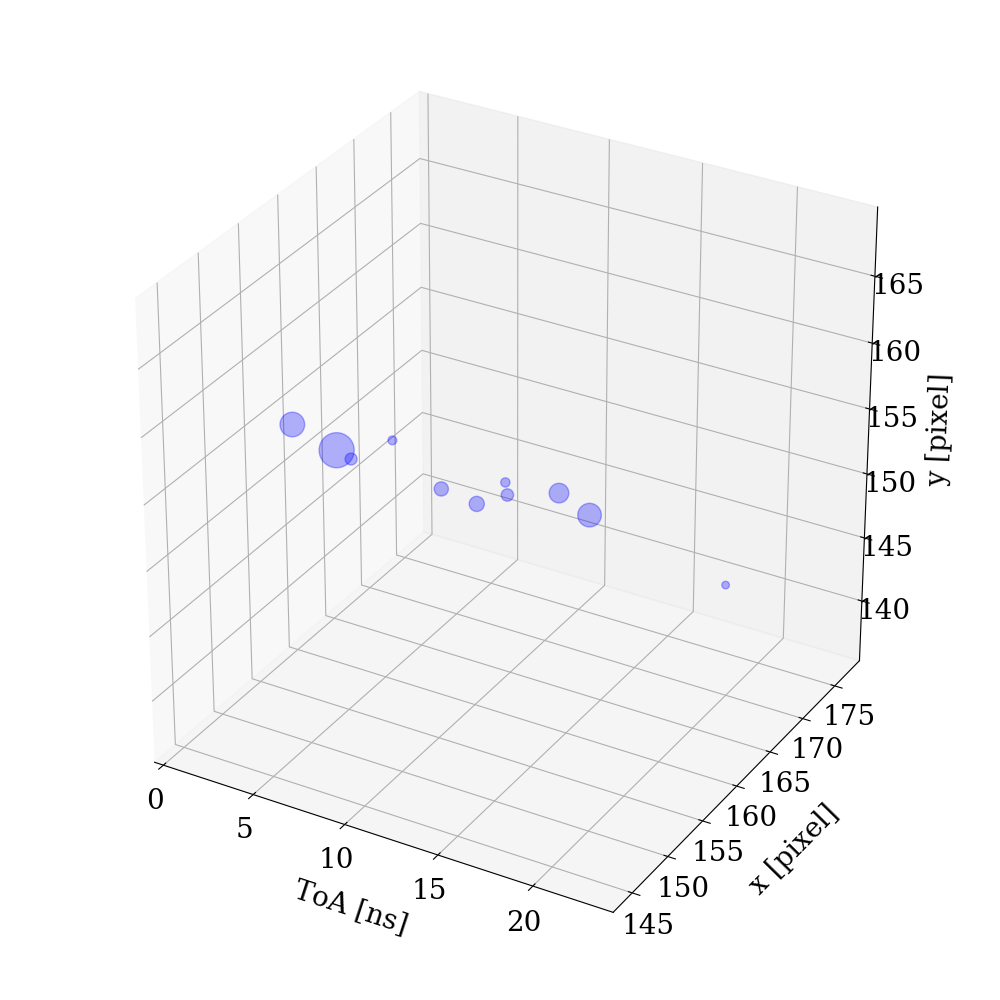}
    \includegraphics[width=.49\linewidth]{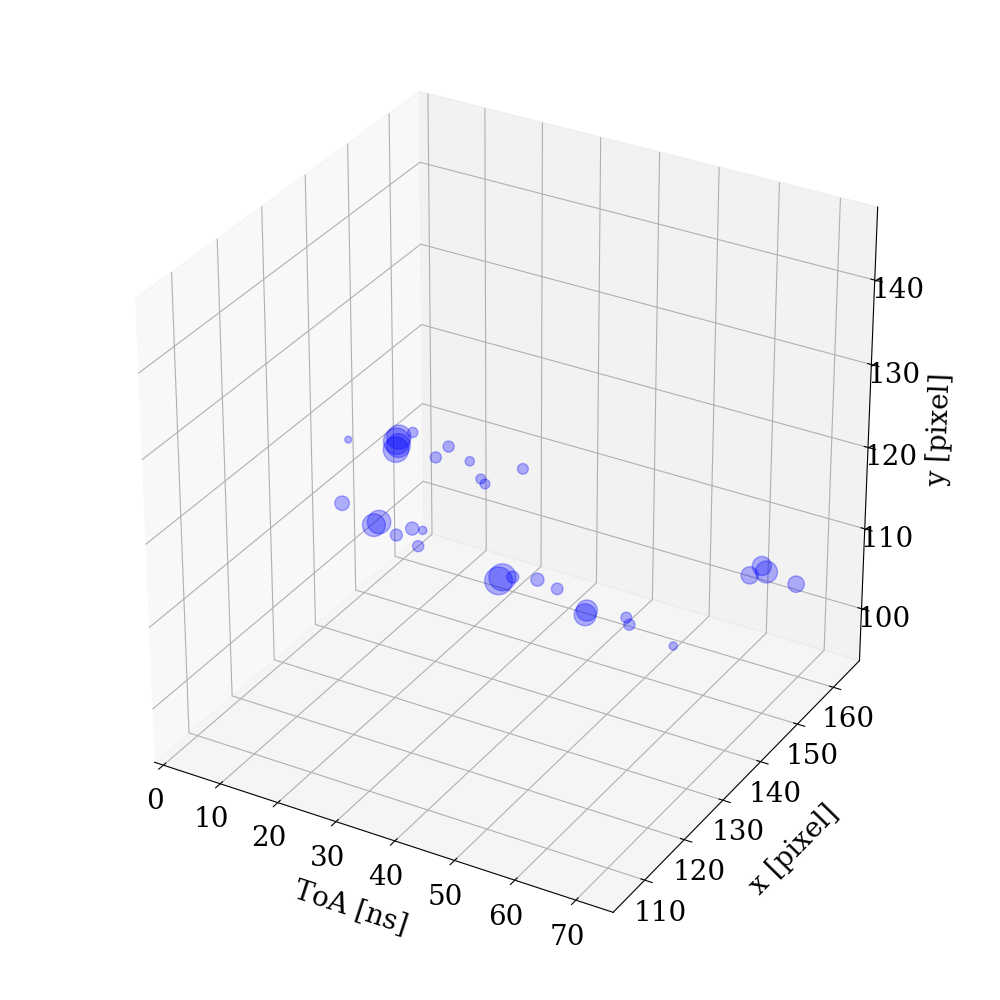}
    \caption{Three-dimensional distributions of the $x$ and $y$ coordinates of hits and their corresponding ToA for hits in single-cluster events (left) and multi-cluster events (right). The ToA information is corrected to minimize the time-walk effect. The size of each dot represents the magnitude of the ToT of the hit on the sensor.}
    \label{fig:types_of_events}
\end{figure}

The clustering algorithm that is presented above relies purely on the timing information of the hits and ignores their spatial information in the grouping of hit pixels into clusters. In addition, it does not account for the effect of photon blending in the images. Since this blending of single photon hits is a characteristic feature of this detection technique, an improved algorithm was developed. The new algorithm reconstructs clusters of multiple hits by using both temporal and spatial information.
Photon hits are first grouped in temporal slices, or "events". Events are reconstructed as collections of four or more temporally close hits, so that no two consecutive hits are registered more than 20~ns apart. The ToA measurements with time-walk correction are used.
The loose requirement on the number of hits is sufficient to remove spurious counts originating from environmental noise, cosmics, and overly sensitive pixels. 
Events are further divided into "clusters", which are defined as groups of at least four hits within a distance of 3~pixels (corresponding to a radius of 3 $\times$ 55~$\mu$m = 165~$\mu$m) or less from the cluster center. The center of the cluster is computed twice with a non-weighted center-of-mass method to minimize the exclusion of outlier hits. Again, no two consecutive hits in a cluster can be registered more than 20~ns apart. 
This combination of spatial and temporal requirements allows to better reconstruct single photons where the standard post-processing algorithm, with no spatial requirements, would blend them. Figure \ref{fig:types_of_events} shows examples of events with a single cluster (left) and multiple reconstructed clusters (right).

\begin{figure}[!htb]
    \centering
    \includegraphics[width=\linewidth]{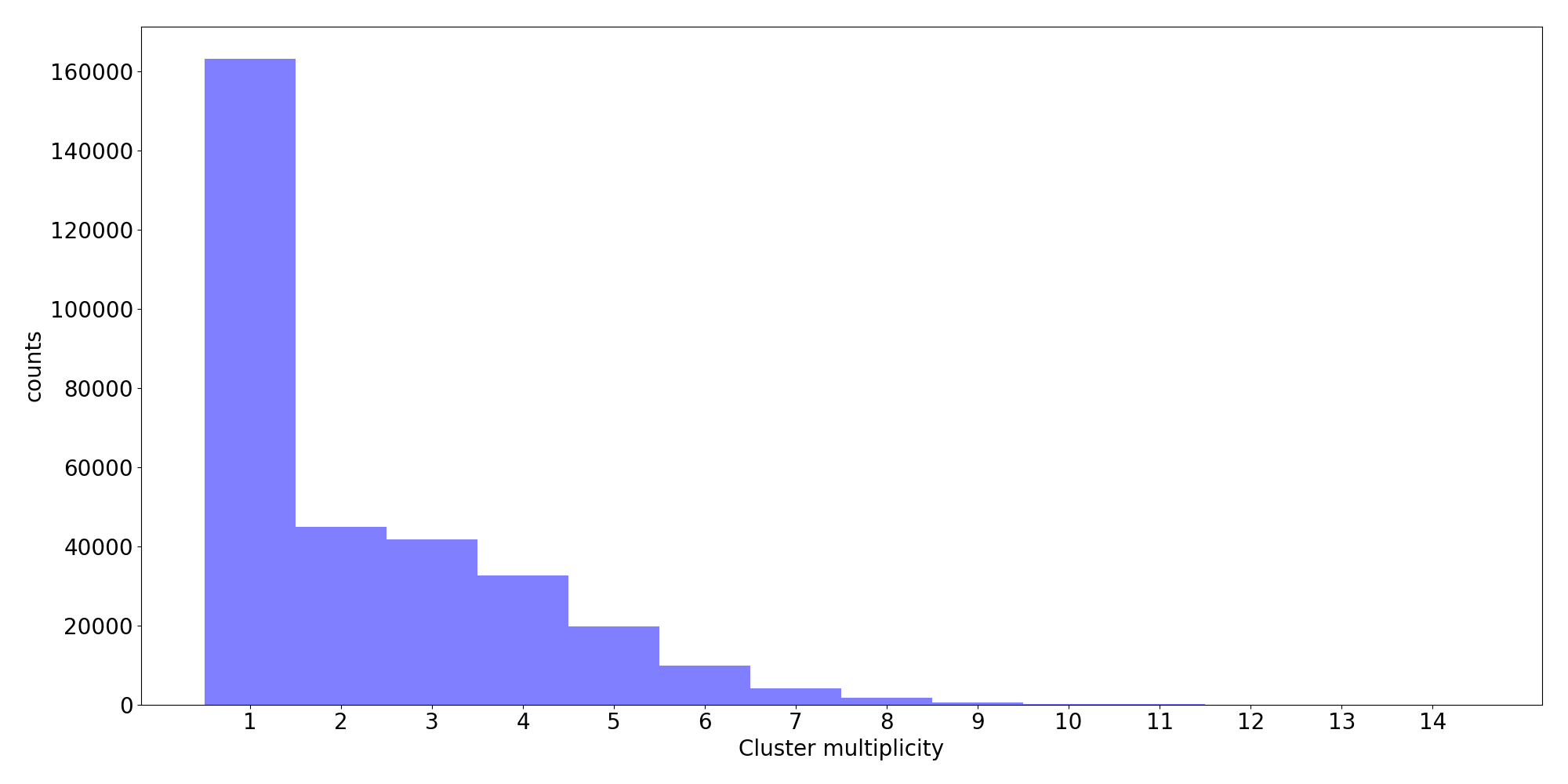}
    \caption{Number of clusters reconstructed in each event by the optimised post-processing algorithm. 
    }
    \label{fig:cluster_multiplicity}
\end{figure}

Being the cluster reconstruction algorithm sensitive to the relative spatial and temporal coordinates of the hits, multiple procedures to remove spurious counts have been exploited. 
The presence of rapidly firing pixels, or "hot pixels", can cause issues to the reconstruction procedure, as it may create spurious clusters of hits very close in time. Pixels are individually masked off based on their firing rate compared to that of their neighbors. A pixel is masked off and therefore not considered in the post-processing reconstruction if none of its neighbors shows a comparable firing rate. This filter is an additional protection against "hot pixels" that are not identified and masked off during the camera calibration procedure. Furthermore, events with a distance of more than 50 pixels (corresponding to a radius of 2.75~mm) from the center of the lens focal point are discarded to remove reflections from the sides of the LYSO scintillator.
Hits with small ToT values have the largest time-walk corrections with considerable uncertainties. To mitigate this effect and improve the timing resolution, hits inside a cluster are discarded if their ToT is less than 20$\%$ of the highest ToT in the cluster.

The number of single-cluster and multi-cluster events can be confronted by looking at Figure \ref{fig:cluster_multiplicity}, which shows the distribution of the number of clusters in each event. Clearly distinguishable is the peak in the distribution around the value of 1, representing events with a single reconstructed cluster, and a long tail at higher values of cluster multiplicity, which corresponds to multi-cluster events.
Single-cluster events are likely to be originated from the $^{241}$Am 59~keV x-ray emission, while multi-cluster events are expected to originate from the interaction of $\upalpha$-particles emitted by $^{241}$Am with the LYSO scintillator.
These two populations are expected to correspond to the respective productions of x-ray photons and $\upalpha$-particles in $^{241}$Am, once the reconstruction efficiencies of the algorithm are taken into account for both production mechanisms. 

\begin{figure}[!htb]
    \centering
    \includegraphics[width=.49\linewidth]{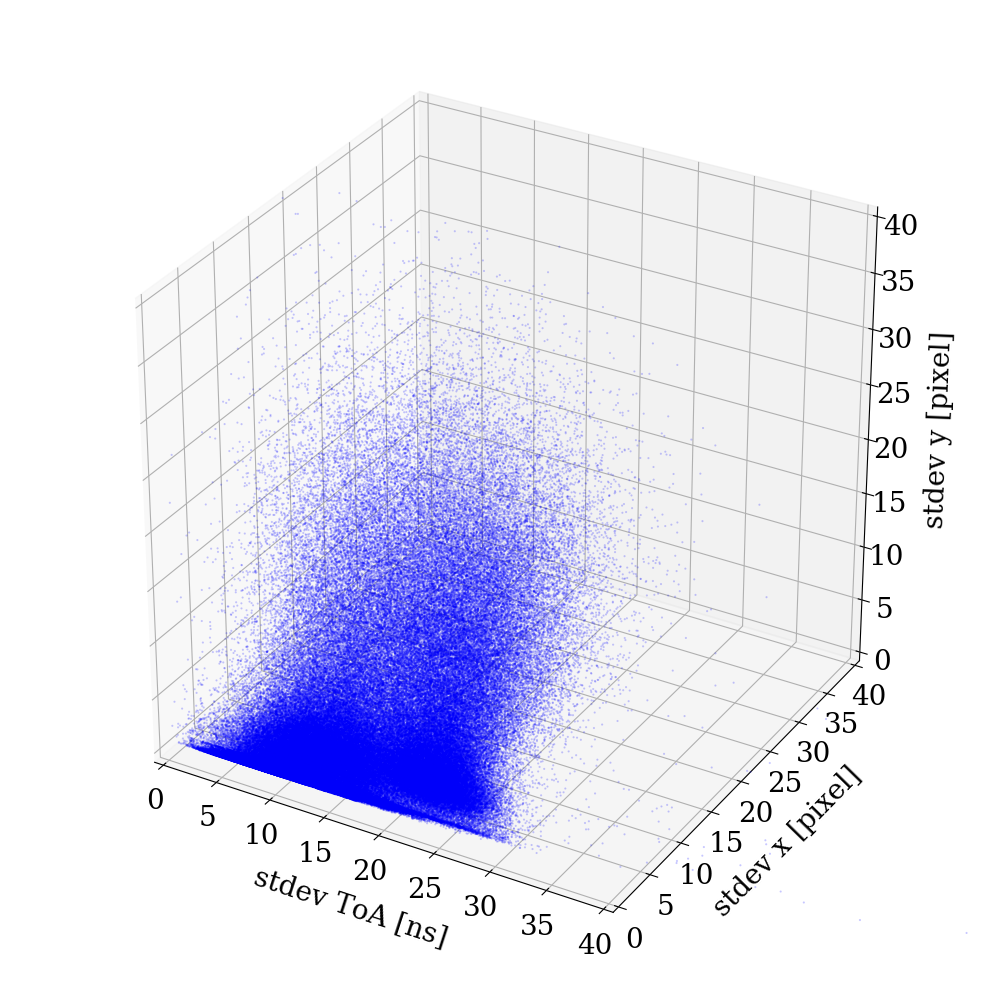}
    \includegraphics[width=.49\linewidth]{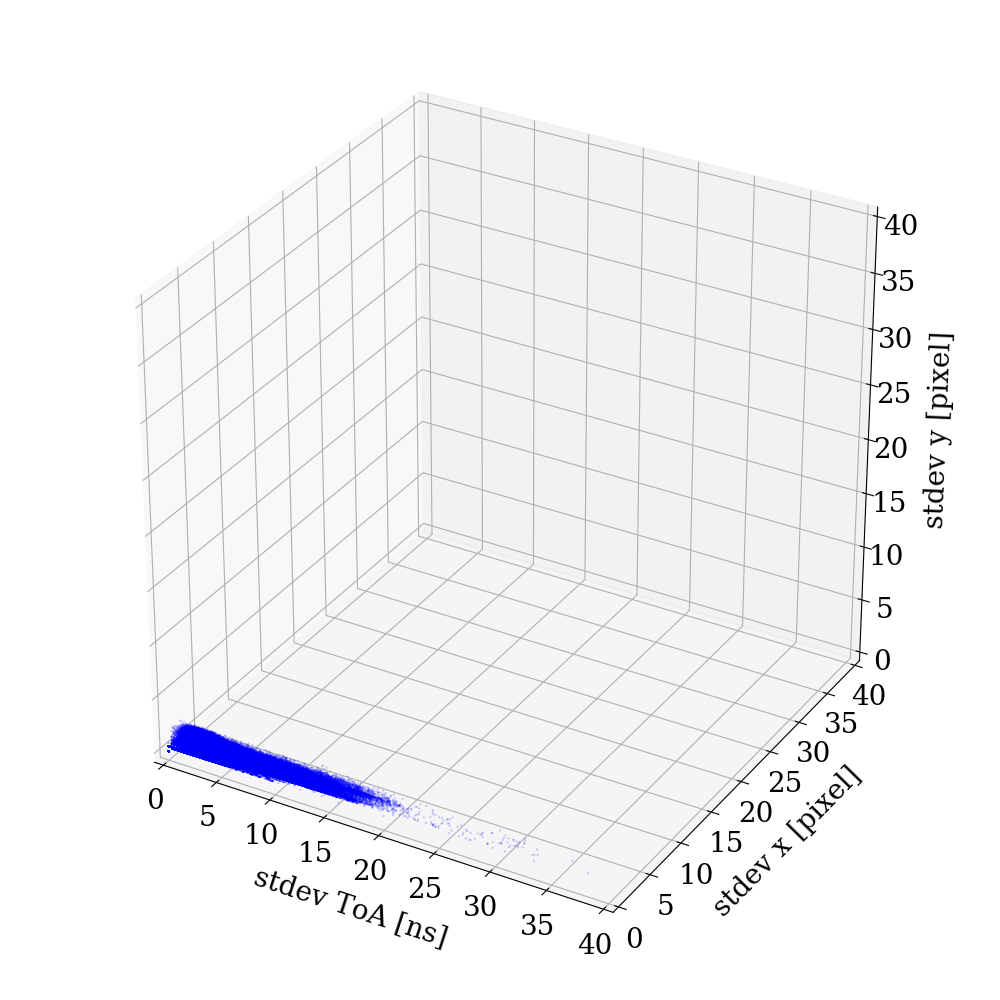}
    \caption{Three-dimensional distributions of the standard deviation of the time-walk-corrected ToA and the corresponding standard deviations of $x$ and $y$ coordinates of all hit pixels in events (left) and clusters (right).}
    \label{fig:divergence}
\end{figure}

While single clusters are expected to be confined in space, events composed by multiple clusters may be characterized by large variances in both space and time dimensions. The temporal spreads of clusters and events have similar magnitude and are mostly derived by the decay time of the LYSO scintillator. Figure \ref{fig:divergence} shows the three-dimensional distribution of the standard deviations for the $x$ and $y$ coordinates of all pixels in events (left) and clusters (right) along with the corresponding standard deviation of ToA using the optimised reconstruction algorithm. The large number of events with small variances is composed by a single cluster, while events with larger divergence tends to have high cluster multiplicity. This result can be compared to the distribution presented in Figure \ref{fig:timecluster}.

\begin{figure}[!htbp]
    \centering
    \includegraphics[width=.8\linewidth]{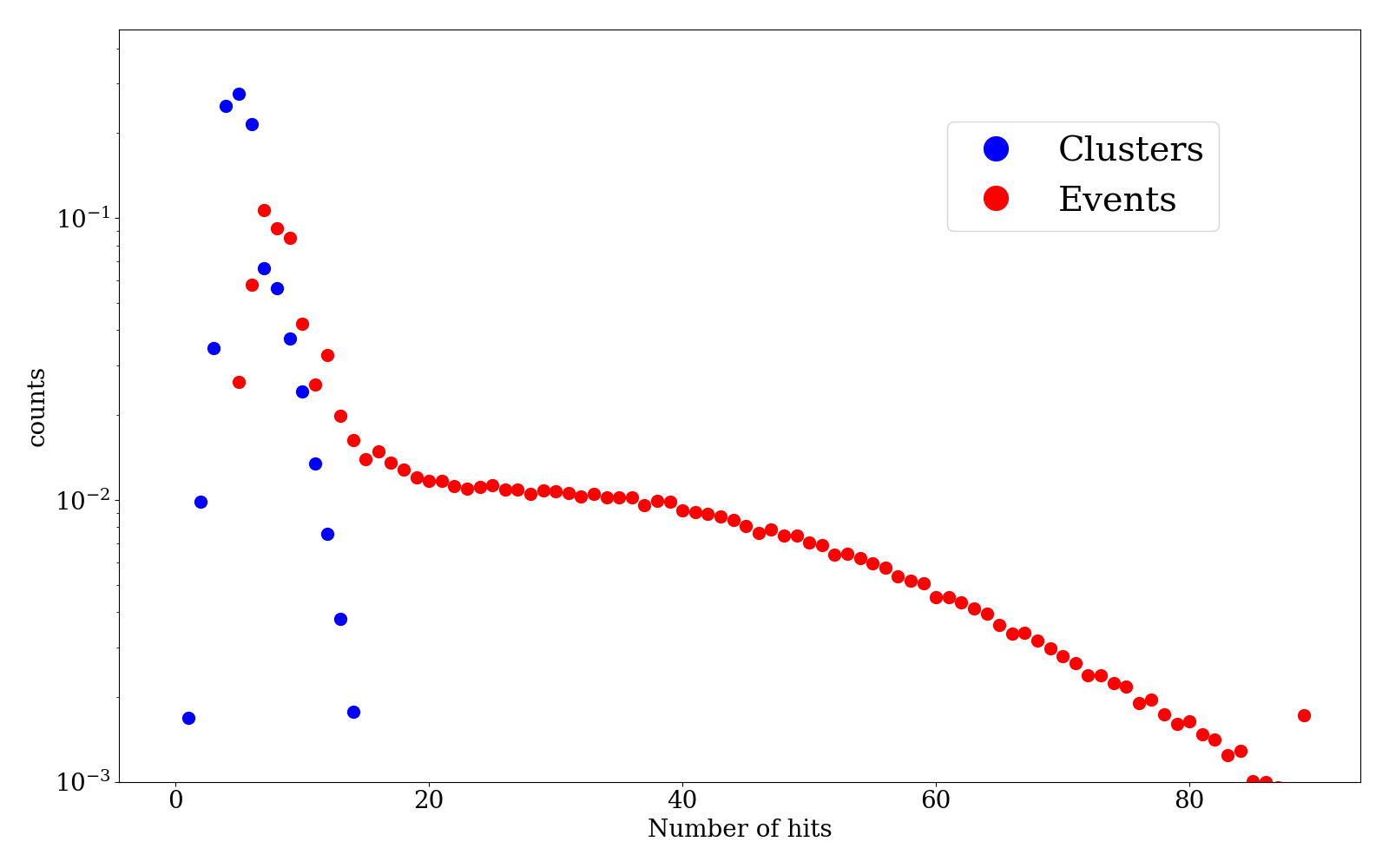}\\
    \includegraphics[width=.8\linewidth]{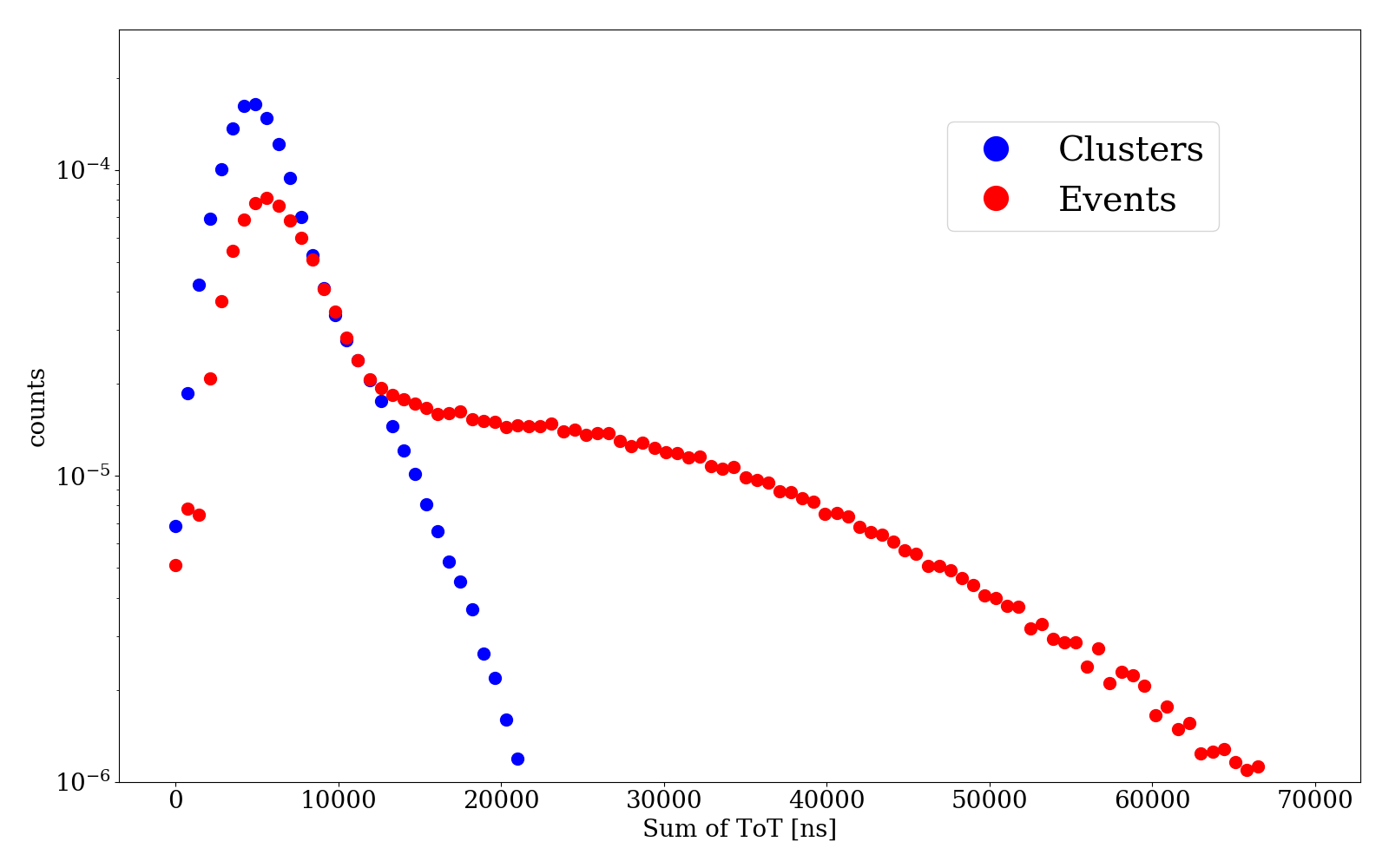}
    \caption{Distributions of number of hits (top) and total ToT (bottom) in an event (red) and in clusters (blue), as reconstructed by the optimised post-processing algorithm. The peaks of the distributions associated to reconstructed events are compatible to those of single-photon (cluster) events, corroborating the hypothesis that those regions are populated primarily by single-photon events, while the tails at higher values are populated by multi-cluster events. The distributions are shown in logarithmic scale on the y axis.
    }
    \label{fig:hits_multiplicity}
\end{figure}


Finally, the presence of these two distinct populations is directly reflected into the composition of the cluster and events at hit level. Figures \ref{fig:hits_multiplicity} shows the distributions of number of hits (left) and total ToT (right) of each event and cluster. The total ToT is defined as the sum of the ToT values associated to all the hits in an event or a cluster. Both plots show a peak in the event distribution compatible to that of the cluster distribution. This observation corroborates the hypothesis that the peak regions of the distributions of the number of hits and total ToT are populated primarily by single-photon events, while the tails at higher values of these observables are populated by multi-cluster events.



\section{Discussion}
\label{sec:discussion}

Section \ref{sec:analysis} presented a novel technique for imaging of $\upalpha$-particles using a fast optical camera and showed that the $\upalpha$-particles can be detected with temporal resolution of about 10 ns and can be resolved spatially. This section presents an estimate of the signal produced in these experiments by 5.49~MeV $\upalpha$-particles from an $^{241}$Am source in a LYSO scintillator. 

The light yield in LYSO is equal to 27,600 photons/MeV \cite{lyso}, resulting in about 152,000 photons produced in the scintillator by 5.49~MeV $\upalpha$-particles. For a correct estimate, quenching effects occurring during a compact release of the $\upalpha$-particle energy \cite{Ogawa2018, vonKrosigk2016, YANAGIDA2018, KOBA2011} need to be accounted for as they could considerably suppress the light yield. A suppression factor of approximately 11 can be directly deduced from the results in Ref.~\cite{AHMADOV2013}, which decreases the light yield to 13,800 photons. 

After leaving the scintillator, the light direction changes in accordance with Snell's law, due to different refraction indices of LYSO and air, correspondingly 1.82 and 1.00 \cite{lyso} respectively. Taking this effect into account in the acceptance calculation, a collection efficiency for the emitted light of about 0.5\% is estimated when considering the 25~mm diameter lens and 100~mm distance between the scintillator and input window of the intensifier. This corresponds to 69 collected photons, which will be registered with about 20\% photon detection efficiency (PDE), resulting in $\approx14$ photon hits recorded by the camera. This estimate is an upper limit, since reflections and other sources of optical attenuation are not accounted for; nevertheless it is close to the observed number of photons in the recorded $\upalpha$-particle events. It is to be noted that Figure \ref{fig:cluster_multiplicity} could be interpreted as a distribution of the number of detected photons, however it is severely affected by the blending, which reduces the number of reconstructed clusters.

The ionization cloud produced by the $\upalpha$-particles is very compact in all directions and is produced within 10 micron from the surface of the scintillator. For optimal light collection, the lens should be focused on that surface. The performance will have a weak geometrical dependence on the thickness of the scintillator since the collection angle will be slowly increasing with the thickness due to the refraction happening closer to the lens. The light collection efficiency would increase by about 2\% if the scintillator thickness changes from 0.5 to 1~mm.

Optical signal from the 59~keV x-rays produced by the same source will not suffer from the quenching. However the total light yield from the x-rays is considerably smaller than that from the $\upalpha$-particles. After accounting for the same efficiency factors, the average number of photons per x-ray registered in the camera is estimated to be equal to 1.6. 

Emission of photons will follow the distribution of the scintillator decay time, about 40 ns, and so, statistically, the timing resolution can be crudely estimated as the decay time divided by the square root of the registered photon number, hence about 10 ns for the tested configuration.

As already discussed, the proposed optical technique could be used for registration of x-rays or, in fact, for all other types of ionizing radiation, which would produce flashes of light in a scintillator. For the x-rays, the registered signal in the tested configuration is expected to be on average in the range of $0.3 - 3$ single photons per x-ray for the corresponding x-ray energy range of $10 - 100$~keV. It must be noted that the conventional solid-state x-ray detectors may have limited sensitivity in this energy range, especially towards its higher energy bound and the optical approach could have advantages.

Since some of the most important detection techniques for neutrons, such as boron-10 based, result in emission of $\upalpha$-particles, the proposed approach may also find applications in neutron detection.
X-rays, charged particles and neutrons could penetrate the full thickness of the scintillator, so the interaction point can be at any depth. In this case it is beneficial to have the scintillator as thin as practical to preserve the sensitivity and spatial information. As the lens is focused at a certain depth in the scintillator, any light produced elsewhere will be out of focus and may not be efficiently collected. 

A clear advantage of this technique is the free-space light collection, which can be performed remotely from large distances and is very flexible. The optical approach could allow a large field of view covered with a single camera with appropriate lens with a demagnification factor. Obviously, the light collection will be inversely proportional to the square of distance, so for the tested configuration the expected number of photons will be equal to one for a 36~cm separation between the scintillator and camera. For larger distances the light collection will be accordingly decreased as well as the $\upalpha$ detection efficiency, which, of course, could become substantially smaller than 100\%. The time resolution would deteriorate too, and, for a number of detected photons much smaller than one, it will be determined by the scintillator decay time. However, for sufficiently strong $\upalpha$-emitters this approach could allow imaging from large distances just by pre-installing scintillators and pointing the fast optical camera towards them  when needed. 

\section{Conclusions}
\label{sec:conclusions}

The viability of a novel technique for imaging of $\upalpha$-particles was proven by employing a fast optical camera with single photon sensitivity. The camera collects and time-stamps photons produced by the particles in a thin layer of LYSO scintillator with nanosecond scale resolution. An optimized reconstruction algorithm was developed to exploit the full spatial and temporal information available for each registered photon in the camera. Similar optical techniques could be applicable to other types of ionizing radiation such as x-rays, charged particles and neutrons.


\acknowledgments

We thank Gabriele Giacomini, Justine Haupt, Boris Shwarts, Jeph Wang, Craig Woody for useful discussions. This work was supported by the U.S. Department of Energy under grant DE-SC0012704, by the BNL Laboratory Directed R$\&$D (LDRD) awards 19-030 and 18-038, as well as by the grant LM2018109 of Ministry of Education, Youth and Sports as well as by Centre of Advanced Applied Sciences CZ.02.1.01/0.0/0.0/16-019/0000778, co-financed by the European Union.


\bibliographystyle{unsrt}
\bibliography{Am241}

\end{document}